\documentclass[preprint2]{aastex}
\usepackage{lscape}

\newcommand{\xmm}{{\em XMM-Newton}}
\newcommand{\chan}{{\em Chandra}}

\shorttitle{X-ray observations of Fermi pulsars}
\shortauthors{Marelli et al.}

\begin{document}

\title{Radio-quiet and radio-loud pulsars: similar in $\gamma$-rays but different in X-rays}

\author{
M. Marelli\altaffilmark{1},
R. P. Mignani\altaffilmark{1,2},
A. De Luca\altaffilmark{1,3},
P. M. Saz Parkinson\altaffilmark{4,5},
D. Salvetti\altaffilmark{1},
P.R. Den Hartog\altaffilmark{6},
M.T. Wolff\altaffilmark{7}
}

\affil{\altaffilmark{1} INAF - Istituto di Astrofisica Spaziale e Fisica Cosmica Milano, via E. Bassini 15, 20133, Milano, Italy}
\affil{\altaffilmark{2} Kepler Institute of Astronomy, University of Zielona G\'ora, Lubuska 2, 65-265, Zielona G\'ora, Poland}
\affil{\altaffilmark{3} Istituto Nazionale di Fisica Nucleare, Sezione di Pavia, Via Bassi 6, I-27100 Pavia, Italy}
\affil{\altaffilmark{4} Santa Cruz Institute for Particle Physics, Department of Physics, University of California at Santa Cruz, Santa Cruz, CA 95064, USA}
\affil{\altaffilmark{5} Department of Physics, The University of Hong Kong, Pokfulam Road, Hong Kong, China}
\affil{\altaffilmark{6} Stanford University HEPL/KIPAC, 452 Lomita Mall, Stanford, CA 94305-4085}
\affil{\altaffilmark{7} Space Science Division, Naval Research Laboratory, Washington, DC 20375-5352, USA}

\begin{abstract}
We present new \chan\ and \xmm\ observations of a sample of eight radio-quiet $\gamma$-ray pulsars detected by the {\em Fermi} Large Area Telescope.
For all eight pulsars we identify the X-ray counterpart, based on the X-ray source
localization and the best position obtained from $\gamma$-ray pulsar timing.
For PSR\, J2030+4415 we found evidence for a $\sim 10\arcsec$-long pulsar wind nebula. Our new results consolidate the work from \citet{mar11} and confirm that, on average,
the $\gamma$-ray--to--X-ray flux ratios (F$_{\gamma}$/F$_X$) of radio-quiet pulsars are higher than for the radio-loud ones. Furthermore, while the F$_{\gamma}$/F$_X$ distribution 
features a single peak for the radio-quiet pulsars, the distribution is more dispersed for the radio-loud ones, possibly showing two peaks. We discuss possible implications of
these different distributions based on current models for pulsar X-ray emission.

\end{abstract}


\section{Introduction}

The launch of the {\em Fermi} $\gamma$-ray Space Telescope in 2008 marked a revolution in pulsar $\gamma$-ray astronomy
\citep[for a recent review see][]{car14}.
The Large Area Telescope \citep[LAT,][]{atw09} onboard {\em Fermi} has detected tens of millisecond and young--to--middle-aged rotation-powered $\gamma$-ray pulsars,
with 117 listed in the Second {\em Fermi} LAT Catalog of $\gamma$-ray pulsars \citep[2PC,][]{abd13}. This number has already risen to more
than 160\footnote{http://tinyurl.com/fermipulsars}. About 30\% of the LAT pulsars are radio-quiet.
In the 2PC, radio-quiet (RQ) pulsars are defined as those that have not been detected in radio down to a flux density limit of S$_{1400}=30\mu$Jy
at 1400 MHz, whereas radio-loud (RL) pulsars are those detected above this limit. Some pulsars that have been detected in radio with a flux density below S$_{1400}=30\mu$Jy  are defined radio faint (RF).
Assuming that radio
emission is produced close to the magnetic poles, this large fraction of
RQ $\gamma$-ray pulsars
suggests that the $\gamma$-ray beam is broader and at a larger angle from the magnetic poles with respect
to the radio beam \citep[see][and references therein]{car14}, thus making its detection less sensitive to geometric effects.
This piece of evidence has
been crucial to verify the predictions of pulsar magnetospheric models, such as the outer gap and the polar gap ones
\citep{che86,har04}. Multi-wavelength studies of $\gamma$-ray pulsars are key to map the geometry
of the different emission regions in the pulsar magnetosphere, investigate possible connections between different emission
processes, and study their efficiencies as a function of energy. For instance, \citet{mar11,mar12}
showed that the distribution of the $\gamma$--to--X-ray flux ratio $F_{\gamma}/$F$_{\rm X}$ is different for RL and
RQ pulsars, being narrower and peaking at higher values for the latter. What is behind the difference in the relative
$\gamma$ and X-ray efficiencies of these two classes of $\gamma$-ray pulsars is not clear yet. While some clues are 
evident from the study of a handful of RQ pulsars detected in X-rays \citep[see e.g.][]{mar13,mar14a,mar14b}, studying a larger sample 
allows one to explore differences across the parameter space.
Furthermore, the detection of RQ $\gamma$-ray pulsars
in X-rays (and in the optical) can provide information, for instance,
on the pulsar dynamics and distance, traditionally obtained in the radio band.
In fact, for RL pulsars the measurement of the dispersion of the pulses at different radio frequencies allows one to estimate the free electron column density,
from which the distance to the pulsar is obtained  from a gas distribution model \citep{cor02}. Alternatively, direct distance measurements
for RL pulsars are obtained from the radio pulsar parallaxes.
The measurement of the proper motion and distance of the RQ LAT pulsar PSR\, J0357+0352 with
\chan\ and \xmm\ \citep{del13,mar13} is a spectacular example of distance estimation is possible also for
RQ pulsars and using a different method.

The X-ray satellites that have yielded the highest pulsar detection rates have traditionally been
 \chan\ and \xmm\, due to their spatial, timing and
spectral resolutions, along with their low background rates.
Here, we present the results of new follow-up observations of a sample of RQ LAT pulsars
with no previous X-ray detections. Our observations are summarized in Section
2, while our data analysis and results are described in Section 3. In Section 4 we compare
the X and $\gamma$-ray properties of all RQ LAT pulsars, making some considerations based
on current models of $\gamma$-ray and X-ray emission from pulsars.

\section{Target selection and observation description}

We selected our target pulsars for \chan\ and \xmm\ observations among those RQ 
LAT pulsars that have no,
or uncertain, detection in X-rays (see Table 5 of 2PC). In order to predict the non-thermal X-ray
flux of these pulsars, we relied on the observed $F_{\gamma}/$F$_{\rm X}$ distribution of the RQ pulsars family,
favoring pulsars with relatively high $\gamma$-ray flux and small pseudo-distance D$_{\gamma}$, as inferred from
the comparison between flux and luminosity of LAT pulsars with known distance \citep{saz10}.
Our sample includes eight LAT pulsars, spanning three orders of magnitude in characteristic ages ($\tau_c\equiv P/2\dot{P}$,
where P is the neutron star spin period and $\dot{P}$ is derivative) and two orders
of magnitude in spindown energies ($\dot{E}\equiv 4\pi^2$I$\dot{P}/P^3$ , where the moment of inertia I is assumed to be 10$^{45}$ g cm$^{-2}$). Their $\gamma$-ray timing coordinates,
which we used as a reference for the X-ray counterpart identification, and their timing parameters are summarized
in Table\, 1, together with the values of the pseudo-distance D$_{\gamma}$, as defined in \citet{saz10}. We took the pulsar timing parameters
and the pseudo-distance values from the compilations in \citet{abd13}, whereas 
we took the $\gamma$-ray timing positions and errors from the most recent compilation of the LAT $\gamma$-ray
Pulsar Timing Models \footnote{\url{https://confluence.slac.stanford.edu/display/GLAMCOG/LAT+Gamma-ray+Pulsar+Timing+Models}}.
This compilation reports the results of the analysis of five years of {\em Fermi} data by using a new,
advanced timing analysis pipeline \citep{ker15} that handles glitch detection and fitting in the
presence of timing noise much better than previous codes. We note that before this work, the positional errors
based on $\gamma$-ray timing were correctly calculated (e.g. taking into account timing noise and systematic errors) only for a small number of bright pulsars \citep[][see e.g.]{ray11}.

{\it Swift} already observed all the $\gamma$-ray pulsars in our sample in short ($\approx$ 5--10 ks)
snapshot observations. Through the re-analyses of these data, we found no X-ray source that could be
positionally associated with the pulsar \citep[for a more detailed discussion see][]{mar12}. Three of
these pulsars were also re-observed by {\em Suzaku}: two of them (PSR\, J1429$-$5911 and J1957+5033)
remained undetected, also owing to the short exposure times,
whereas a possible, marginal X-ray detection was obtained for PSR\, J1838$-$0537.

Here, we report on five new 25 ks \chan\ observations of PSRs J1429$-$5911, J1957+5033, J2028+3332, J2030+4415, and
J2139+4716 (observation ids 14825, 14828, 14826, 14827 and 14829, respectively) and 15 ks new \chan\
observations of PSRs J0734$-$1559 and J1846+0919 (observation ids 13792 and 13793, respectively). We observed
all the pulsars with the Advanced CCD Imaging Spectrometer (ACIS). In order to detect the highest number
of counts from our targets we obtained the ACIS-S observations in the VFAINT mode, with
the targets positions placed on the back-illuminated ACIS S3 chip. We observed PSR\, J1838$-$0537 with
\xmm\ for 44 ks (obs. id 0720750201) using The EPIC (European Photon Imaging Camera) PN and the
two Metal Oxide Semi-conductor (MOS) cameras in the Full Frame mode with medium optical filters,
due to the presence of moderately bright stars within their field--of--view (FOV). All the observations
were carried out between 2012 September 16 and 2014 April 15.

\section{Data analysis and results}

For the \chan\ data analysis we used the \chan\ Interactive Analysis of Observation (CIAO) software version 4.3.
We re-calibrated our event data by using the {\tt chandra\_repro} script and selected events in the 0.3--10 keV
energy range. Depending on the total number of counts, we ran the {\tt wavdetect} detection tool in different
energy ranges, taking into account the exposure maps. For each pulsar, we found a single X-ray source within
(or close to) the 95\% $\gamma$-ray timing error box, with a detection significance above 5$\sigma$ and source
counts varying from 12 (PSR\, J1846+0919) to 123 (PSR\, J1957+5033). For each X-ray source we performed a
spectral analysis. We extracted the source counts from a 2\arcsec\ circle radius around the X-ray position and
the background counts from a surrounding annulus, whose radius we chose on a case by case basis in order to
avoid contamination from serendipitous nearby sources. We used the CIAO tool {\tt specextract} to create the
spectra, response matrix and effective area files and analyzed the spectra with XSPEC (version 12.8.1).
We used the C-statistic \citep{cas79}, based on the application of the likelihood ratio and recommended
for cases with low statistics and low background.  As done in \citet{mar13}, we performed a brightness profile analysis on all X-ray sources, revealing no evidence
for extended emission, with the exception of PSR\, J2030+4415 (see Section 3.7).

For the analysis of the 44-ks XMM observation of PSR\, J1838$-$0537, we used the \xmm\ Science Analysis Software (SAS) v13.0.
We performed a standard analysis of high particle background \citep[following][]{del05} and cross-checked the results with the SAS tool {\tt bkgoptrate} (also used for the 3XMM
source catalog\footnote{\url{http://xmmssc-www.star.le.ac.uk/Catalogue/xcat\_public\_3XMM-DR4.html}}), obtaining very good agreement. 
After the subtraction of bad time intervals affected by soft proton X-ray flares,
we obtained a net exposure time of 23 ks. We selected 0-4 pattern events for PN and 0-12
for the MOS detectors in the 0.3--10 keV energy range. From the cleaned events, we ran
the source detection according to two different methods: we used the SAS 
{\tt edetect-chain} and the CIAO {\tt wavdetect} tools. Both of them detected a single
X-ray source within few arcsecs from the pulsar $\gamma$-ray timing position.
The source appears point-like, with its brightness profile consistent with the detector's Point Spread Function.
In order to maximize the source signal--to--noise (S/N), we extracted spectral counts from a 25\arcsec\ radius
circular region around the computed X-ray position and the background from a source-free, nearby, 40\arcsec\ 
radius circle on the same CCD. We generated ad hoc response matrices and effective area files using the
SAS tools {\tt rmfgen} and {\tt arfgen}. To increase the statistics, we added the two MOS spectra by using
the HeaSoft tool {\tt mathpha} and the two response matrices and effective area with {\tt addrmf} and {\tt addarf}.
After background subtraction, we extracted 200 and 176 net source counts from the PN and the two MOS, respectively. Owing to the high background
(64\% and 43\% of the extracted counts from the PN and MOS, respectively), we used the $\chi^{2}$ statistic for the spectral fit.

Due to the low statistics of our X-ray sources, we assumed the pulsar pseudo-distance $D_{\gamma}$ to normalize
the value of the integrated Galactic N$_{\rm H}$ in the pulsar direction, computed according to the recalibration
\citep{sch11} of the extinction maps from \citet{sch98}. Four of our eight sources have fewer
than 20 net counts, which are not adequate to perform detailed spectral fits. Therefore, for these sources we fixed
the power-law (PL) photon index and blackbody (BB) temperature to representative values of 2 and 200 eV, respectively.
Following, e.g., \citet{mar11}, these are the fitted average of the values measured for pulsars detected in the X rays.
Errors in the spectral parameters are reported at the 90\% confidence level (c.l.).

\begin{figure*}
\centering
\includegraphics[angle=0,width=15cm]{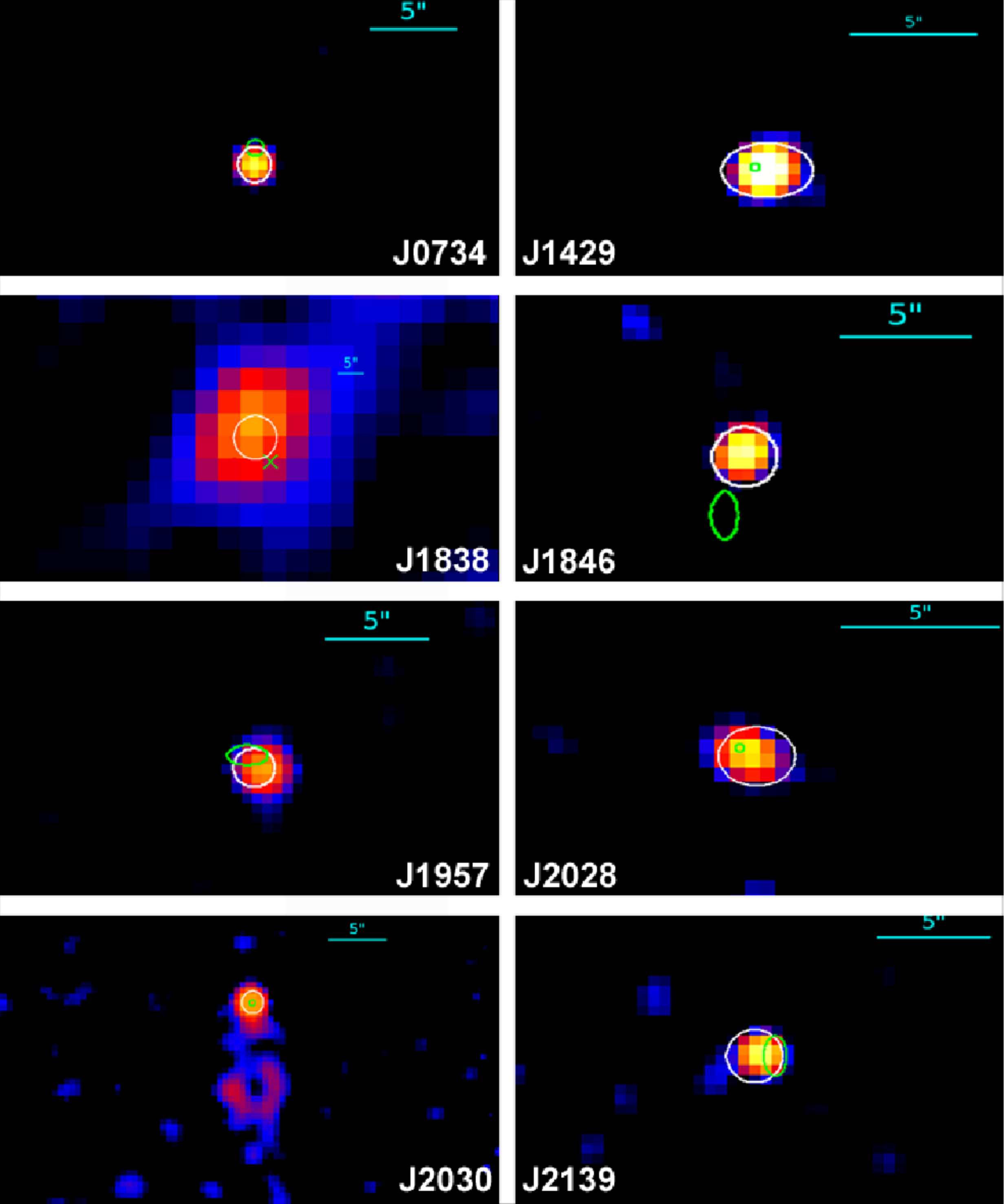}
\protect\caption{{\footnotesize \chan\ and \xmm\ images of the pulsars listed in Table\, 1, taken in the 0.3--10 keV energy range. For a better visualisation, we applied a Gaussian filter with a kernel radius of 3\arcsec. In each panel, the best-fit X-ray position (90\% confidence errors) is shown by the white circle and the LAT $\gamma$-ray timing position (1$\sigma$ errors) by the green ellipses. The latter are taken from the most recent compilation of the LAT $\gamma$-ray Pulsar Timing Models. The position of PSR\, J1838$-$0537 is taken from \citet{ple12a}, where the positional error is not reported, and is marked by the green cross.}
\label{ima}}
\end{figure*}

The errors on the X-ray positions reported in the following sub-sections are purely statistical, at a $3 \sigma$
confidence. To them, we have to add the 90\% c.l. systematical errors associated with the absolute accuracy of
the satellite aspect solution, which are 0\farcs8 and 1\farcs5 per coordinate
for \chan\footnote{{\texttt http://cxc.harvard.edu/cal/ASPECT/celmon/}} and \xmm\footnote{calibration technical note XMM-SOC-CAL-TN-0018}, 
respectively. According to the logN-logS distribution of \chan\ Galactic sources \citep{nov09},
we can estimate the probability of a chance detection of an
X-ray source within a representative LAT $\gamma$-ray timing error box (1\arcsec\ $^2$, see Figure \ref{ima}), with X-ray flux similar or
greater than the measured ones, to be about 3$\times$10$^{-5}$. Thus, based on positional
coincidence, we consider our identifications to be secure. Figure \ref{ima} shows the positions and
errors of $\gamma$-ray pulsars and the associated X-ray counterparts. The computed X-ray spectral
parameters, unabsorbed X-ray fluxes, and $\gamma$--to--X-ray flux ratios of our eight pulsars are summarized in Table 2.

\subsection{PSR\, J0734$-$1559} 

PSR\, J0734$-$1559 (P=156 ms) was identified as a middle-aged (0.2 Myr) $\gamma$-ray pulsar during a blind search for pulsations
from the unidentified LAT source 1FGL\, J0734.7$-$1557 \citep{saz11}. We detected its X-ray counterpart
at $\alpha_{\rm X} =07^{\rm h} 34^{\rm m} 45\fs7$ ($\pm$0\farcs15);
$\delta_{\rm X} = -15^\circ 59\arcmin 19\farcs8$ ($\pm$0\farcs26), consistent with the pulsar $\gamma$-ray timing coordinates.
The X-ray source is detected with a significance of 8.3 $\sigma$, as computed by {\tt wavdetect}, with 19 net counts.
As described above, we assumed the pulsar pseudo-distance D$_{\gamma}$=1.3 kpc to estimate a N$_{\rm H} =2 \times 10 ^{21}$ cm$^{-2}$.
Keeping these values fixed, we fitted the X-ray spectrum of the pulsar with either a single power law (PL) or blackbody (BB) model.
Due to the low statistics, we also fixed the PL photon index to 2 and the BB temperature to 200 eV.
The best-fit with a BB gives a radius of  the emitting region on the neutron star surface of $76^{+16}_{-11}$ m, computed
for the assumed pseudo-distance D$_{\gamma}$. For the PL model, the unabsorbed X-ray flux in the 0.3--10 keV energy range
is $F^{\rm PL}_{\rm X} = 1.5\pm0.6 \times 10^{-14}$ erg cm$^{-2}$ s$^{-1}$, which gives a $\gamma$--to--X-ray flux ratio
$F_{\gamma}/F_{\rm X} \sim3700$. The BB model, on the other hand, gives an X-ray flux $F^{\rm BB}_{\rm X}=1.2\pm0.4 \times 10^{-14}$
erg cm$^{-2}$ s$^{-1}$ and an $F_{\gamma}/F_{\rm X}>4700$, assuming $F^{\rm BB}_{\rm X}$ as an upper limit on the non-thermal X-ray
flux from the pulsar.

\subsection{PSR\, J1429$-$5911} 

The $\gamma$-ray pulsar PSR\, J1429$-$5911 (P=115 ms) was one of the very first discovered by applying
the blind-search technique \citep{saz10}.
We detected its X-ray counterpart at
$\alpha_{\rm X} =14^{\rm h} 29^{\rm m} 58\fs5$ ($\pm$1\farcs03); $\delta_{\rm X} = -59^\circ 11\arcmin 36\farcs2$ ($\pm$0\farcs45),
with a significance of $8.5 \sigma$ (24 net source counts). As we did in the previous section, we assumed the pseudo-distance
D$_{\gamma}$=1.7 kpc to estimate an N$_{\rm H} =3 \times 10 ^{21}$ cm$^{-2}$ and we kept it fixed in the X-ray spectral fit.
In this way, a fit with a PL gives a photon index $\Gamma_{\rm X}= -0.1\pm0.7$ and
an unabsorbed X-ray flux $F^{\rm PL}_{\rm X} = (3.3\pm1.8) \times 10^{-14}$ erg cm$^{-2}$ s$^{-1}$.
This corresponds to $F_{\gamma}/F_{\rm X} \sim2400$. A fit to the data with a single BB model did
not yield an acceptable temperature (T$>$0.8 keV).

\subsection{PSR\, J1838$-$0537} 

The $\gamma$-ray pulsar (P=145 ms) PSR\, J1838$-$0537 was discovered through a blind search for pulsations
of the unassociated LAT source 2FGL\, J1839.0$-$0539 \citep{ple12a}.
The pulsar is the youngest (5 kyr) and most energetic ($\dot{E} \sim 5.9 \times 10^{36}$ erg cm$^{-2}$ s$^{-1}$) in our sample. 
PSR\, J1838$-$0537 is spatially coincident with the TeV source HESS\, J1841$-$055, hence possibly
associated with a PWN detected at very high energies. A 41.1 ks {\it Suzaku} observation revealed
a candidate X-ray counterpart \citep{ple12a}. However, the non-negligible
probability of a finding a spurious source within the large {\em Suzaku} error circle
\citep[$\approx$ 19\arcsec\ radius][]{uch08}, as well as the low significance of detection
($\sim3\sigma$) made the identification of the X-ray source with the pulsar uncertain.
Owing to the better angular resolution of \xmm\, we detected the pulsar X-ray counterpart at
$\alpha_{\rm X} =18^{\rm h} 38^{\rm m} 56\fs2$ ($\pm$2\farcs65);
$\delta_{\rm X} = -05^\circ 37\arcmin 04\farcs5$ ($\pm$2\farcs69), with a significance of 20.7
$\sigma$ ($\sim 270$ counts; pn+MOS). The X-ray spectrum is best fitted by an absorbed
PL with a photon index $\Gamma_{\rm X} = 0.8^{+1.1}_{-0.9}$ and a fitted
N$_{\rm H}= 2.7^{+4.4}_{-2.1} \times 10^{22}$ cm$^{-2}$ (90\% c.l.). The unabsorbed X-ray flux is
$F^{\rm PL}_{\rm X} = (7.2\pm0.9)\times10^{-14}$ erg cm$^{-2}$ s$^{-1}$, which gives
a $\gamma$-to-X-ray flux ratio $F_{\gamma}$/F$_{\rm X} \sim$ 2600. A fit with a BB spectrum did not yield an acceptable temperature (T$>$1.5 keV).

\subsection{PSR\, J1846+0919} 

PSR\, J1846+0919 (P=225 ms) was identified as a $\gamma$-ray pulsar by \citet{saz10}.
We detected the X-ray counterpart at $\alpha_{\rm X} =18^{\rm h} 46^{\rm m} 25\fs8$ ($\pm$0\farcs41);
$\delta_{\rm X} = 09^\circ 19\arcmin 49\farcs8$ ($\pm$0\farcs45) with a significance of 5.4$\sigma$
(12 net source counts). As usual, from the pseudo-distance we estimated an N$_{\rm H} =2 \times 10 ^{21}$ cm$^{-2}$
and kept it fixed in the X-ray spectral fit. 
A fit with a PL with fixed photon index ($\Gamma_{\rm X}=2$) gives an unabsorbed X-ray flux
$F^{PL}_{\rm X} = (8.7\pm4.8) \times 10^{-15}$ erg cm$^{-2}$ s$^{-1}$, that corresponds to $F_{\gamma}/F_{\rm X} \sim2800$.
Owing to the low statistics, the X-ray spectrum of this source can be fitted also by a BB. From the pseudo-distance
of 1.4 kpc we obtained an emitting radius of 75$_{-18}^{+15}$ m and the unabsorbed flux is
$F^{BB}_{\rm X} = (4.5\pm3.3) \times 10^{-15}$ erg cm$^{-2}$ s$^{-1}$. Assuming, as before,
$F^{\rm BB}_{\rm X}$ as an upper limit to the non-thermal X-ray flux of the pulsar, the lower limit on $F_{\gamma}/F_{\rm X}$ 
would be 5300.

\subsection{PSR\, J1957+5033} 

Like PSR\, J1429$-$5911 and J1846+0919, PSR\, J1957+5033 (P=374 ms) is one of the very first
pulsars discovered by a blind search by \citet{saz10}. We detected the X-ray
counterpart at $\alpha_{\rm X} =19^{\rm h} 57^{\rm m} 38\fs4$ ($\pm$0\farcs20); $\delta_{\rm X} = 50^\circ 33\arcmin 20\farcs8$ ($\pm$0\farcs17).
The pulsar X-ray counterpart is detected with a significance of $29.0 \sigma$ (123 net source counts).
Owing to the adequate statistics, we left the absorption column as a free parameter in our
spectral fits, as in the case of PSR\, J1838$-$0537 (Section 3.3). The X-ray spectrum is best
fit by a PL, which gives an N$_{\rm H} <2.5 \times 10 ^{20}$ cm$^{-2}$, a photon index
$\Gamma_{\rm X} = 2.1\pm0.3$, and an unabsorbed flux $F^{PL}_{\rm X} = (3.0\pm0.5) \times 10^{-14}$
erg cm$^{-2}$ s$^{-1}$. This corresponds to a quite low $F_{\gamma}/F_{\rm X} \sim870$, indeed the lowest among the pulsars in our sample.
A fit with a single BB component is not acceptable (null hypotesis probability of 4$\times10^{-8}$).

\subsection{PSR\, J2028+3332} 

This pulsar (P=176 ms) is one of the nine $\gamma$-ray pulsars discovered by \citet{ple12b}
using a novel blind search technique. We identified the
PSR\, J2028+3332 X-ray counterpart at $\alpha_{\rm X} =20^{\rm h} 28^{\rm m} 19\fs8$ ($\pm$0\farcs45);
$\delta_{\rm X} = 33^\circ 32\arcmin 04\farcs1$ ($\pm$0\farcs23)
with a significance of 5.1 $\sigma$
(15 net source counts). As we did in previous cases, we used the pseudo-distance D$_{\gamma}$=0.9 kpc to derive
an N$_{\rm H} =2 \times 10 ^{21}$ cm$^{-2}$. The fit with a PL with fixed photon index ($\Gamma_{\rm X}=2$) gives an
unabsorbed X-ray flux $F^{\rm PL}_{\rm X} =(5.3\pm3.3) \times 10^{-15}$ erg cm$^{-2}$ s$^{-1}$, and $F_{\gamma}/F_{\rm X}
\sim 10900$, the highest among the pulsars in our sample. We also tried a fit with a BB, which gives an
emitting radius of only $34_{-8}^{+12}$ m, an X-ray flux $F^{\rm BB}_{\rm X}= (2.2\pm1.5)
\times 10^{-15}$ erg cm$^{-2}$ s$^{-1}$, and an $F_{\gamma}/F_{\rm X}>26400$.

\subsection{PSR\, J2030+4415}

The $\gamma$-ray pulsar J2030+4415 (P=227 ms) was also discovered by \citet{ple12b}. The pulsar X-ray counterpart,
at $\alpha_{\rm X} =20^{\rm h} 30^{\rm m} 51\fs4$ ($\pm$0\farcs15); $\delta_{\rm X} = 44^\circ 15\arcmin 38\farcs8$ ($\pm$0\farcs16)
is detected with a significance of 15$\sigma$ (54 net source counts). The 0.3--10 keV X-ray spectrum is described by a PL
with with photon index $\Gamma_{\rm X,PSR}=2.4^{+0.8}_{-0.6}$, for an N$_{\rm H}=6.1^{15.4}_{-6.1} \times 10^{20}$ cm$^{-2}$, which
gives an unabsorbed X-ray flux $F_{\rm X,PSR}=(2.1\pm0.6) \times 10^{-14}$ erg cm$^{-2}$ s$^{-1}$. The low value of the
absorption column (N$_{\rm H} \la 3 \times 10^{21}$ cm$^{-2}$) with respect to the Galactic N$_{\rm H}$ in the pulsar
direction ($\sim 10^{22}$ cm$^{-2}$) agrees with the pseudo-distance D$_{\gamma}$=800 pc.
The $\gamma$--to--X-ray flux ratio for PSR\, J2030+4415 is $F_{\gamma}$/F$_{\rm X} \sim$ 2800.
A fit with a single BB component is not acceptable (null hypothesis probability of 6$\times10^{-9}$).
We found a $\sim 10 \arcsec$-long extended emission around the pulsar, elongated North to South.
To search for extended emission, we applied the CIAO tool {\tt vtpdetect} on an event list purged
from the point-like sources we found in the field using {\tt wavdetect}. This tool, based on the Voronoi Tessellation and Percolation \citep[VTP, see e.g.][]{bos02},
is particularly indicated for the search of extended sources.
This resulted in a false source probability (i.e. the probability that the detection is associated to a real source) for PSR\, J2030+4415
nebula of 10$^{-26}$ and a best-fitting elliptical region with a major axis of $\sim 10 \arcsec$.
Assuming the pseudo-distance of 800 pc, the $"$tail$"$ would have a physical dimension of $\sim$ 0.07 pc.
Since PSR\, J2030+4415 is RQ and \chan\ has observed it only once, we have no proper motion information
yet, thus we cannot say whether the X-ray nebula is aligned with the motion of the pulsar,
although this would be the most likely interpretation. We extracted the net counts from the PWN from an elliptical region of 10\arcsec\ semi-major axis after masking the pulsar.
We computed the background from nearby, source-free elliptical regions. The tail spectrum is well described by a PL,
with $\Gamma_{\rm X, PWN}=1.2^{+0.5}_{-0.4}$,
with the unabsorbed X-ray flux of $F_{\rm X, PWN}=(4.0\pm1.6)\times 10^{-14}$ erg cm$^{-2}$ s$^{-1}$, about twice as large as the pulsar.

\subsection{PSR\, J2139+4716}

This $\gamma$-ray pulsar \citep[P=282 ms;][]{ple12b}
is the oldest (2.5 Myr) and least energetic ($\dot{E} \sim 0.3 \times 10^{34}$ erg cm$^{-2}$ s$^{-1}$) in our sample. 
We detected the pulsar X-ray counterpart at $\alpha_{\rm X} =21^{\rm h} 39^{\rm m} 56\fs0$ ($\pm$0\farcs44);
$\delta_{\rm X} =47^\circ 16\arcmin 13\farcs0$ ($\pm$0\farcs52) with a
significance of 5.5$\sigma$ (16 net source counts). As usual, from the pseudo-distance
D$_{\gamma}$=0.8 kpc we derived N$_{\rm H} =10 ^{21}$ cm$^{-2}$. By keeping it fixed, a fit with a PL
with fixed photon index ($\Gamma_{\rm X}=2$) gives an
unabsorbed X-ray flux $F^{\rm PL}_{\rm X} = 4.7\pm2.5 \times 10^{-15}$
erg cm$^{-2}$ s$^{-1}$, and $F_{\gamma}/F_{\rm X} \sim4900$. The fit with a BB gives a very small emitting
radius of only $33\pm9$ m, a X-ray flux $F^{\rm BB}_{\rm X}= (2.7\pm1.4)
\times 10^{-15}$ erg cm$^{-2}$ s$^{-1}$, and a $F_{\gamma}/F_{\rm X}>8500$.

\subsection{Optical and infrared observations}

None of the pulsars in our sample have been observed in the optical or infrared (IR, 2PC).
For completeness, we also scanned optical/IR data from public imaging surveys.
For PSR\, J1429$-$5911 we found serendipitous IR observations from the VVV \citep[VISTA Variables in the Via Lactea;][]{eme06}
survey, carried out at the ESO's Cerro Paranal Observatory (Chile)
with the 4.1m Visible and Infrared Survey Telescope for Astronomy (VISTA) and an IR camera \citep[VIRCAM;][]{dal06}. 
The fields of PSR\, J1838$-$0537 and PSR\, J1846+0919 were serendipitously observed in the 
UKIDSS \citep[UK Infrared Deep Sky Survey;][]{law07},
carried out at the Mauna Kea Observatory (Hawaii) with the 3.8 m UK
Infrared Telescope (UKIRT) and the Wide Field Camera \citep[WFCAM;][]{cas07}.  
In all these cases, no objects are detected at the \chan\ or \xmm\ positions. The derived upper
limits are orders or magnitudes above the flux levels expected for the pulsar's age and spin-down
power by scaling for the luminosity and distance of other pulsars detected in the IR \citep[e.g.][]{mig12}.
Thus, these results are mainly to be considered as a reference for future deeper follow-up
observations. Our $3\sigma$
limits are Z$\ga$21; Y$\ga$21; J$\ga20$; H$\ga$19; $K\ga19$ (PSR\, J1429$-$5911),
J$\sim$17, H$\sim$19, K$\sim$18 (PSR\, J1838$-$0537), and J$\sim 20.5$, H$\sim 19.4$, K$\sim 18.6$ (PSR\, J1846+0919).

\section{Discussion and conclusions}

Our \chan\ and \xmm\ observations yielded the identification of the X-ray counterparts of eight
RQ LAT pulsars. Thus, the number of RQ LAT pulsars detected in the X-rays amounts now to 28, with only 11 
still missing an X-ray counterpart. This number has to be
compared with the 30 RL LAT pulsars (out of 42) detected in the X-rays.
Our enlarged X-ray database is important for a better
understanding of the multi-wavelength emission properties of LAT pulsars. In particular, focussing on the differences
between young--to--middle-aged RQ and RL pulsars, \citet{mar11} found that, for a given $\gamma$-ray flux,
the former are intrinsically fainter in the X-ray band than the latter. Thus, the distance-independent F$_{\gamma}$/F$_X$
turned out to be a useful parameter to investigate the differences between the X- and $\gamma$-ray emission properties of
the two pulsar families.
For consistency, we followed the same approach as in Marelli et al.\ (2011) and we used our much enlarged sample of
X-ray-detected RQ pulsars to improve their analysis. 

We note that millisecond pulsars (MS) were not included in the analysis of \citet{mar11},
owing to the fundamental differences between their magnetospheres and those of young/middle-aged pulsars \citep[e.g.][]{joh14}.
In fact, owing to their shorter periods, MS pulsars are though to have a smaller magnetosphere than young pulsars,
where the radius of the last closed field line is defined by the so-called light-cylinder radius R$_{LC}$ = P $c$/2$\pi$. Moreover, given their much longer and more complicated evolutionary history,
the magnetic field topology in MS pulsars may be more complex than those of younger
pulsars. This may in part explain why their pulse profiles are more intricate and diverse.
Lastly, MS pulsars generally seem to be more massive than their younger counterparts. See \citet{ven14} for a more detailed discussion.

\begin{figure*}
\centering
\includegraphics[angle=0,width=15cm]{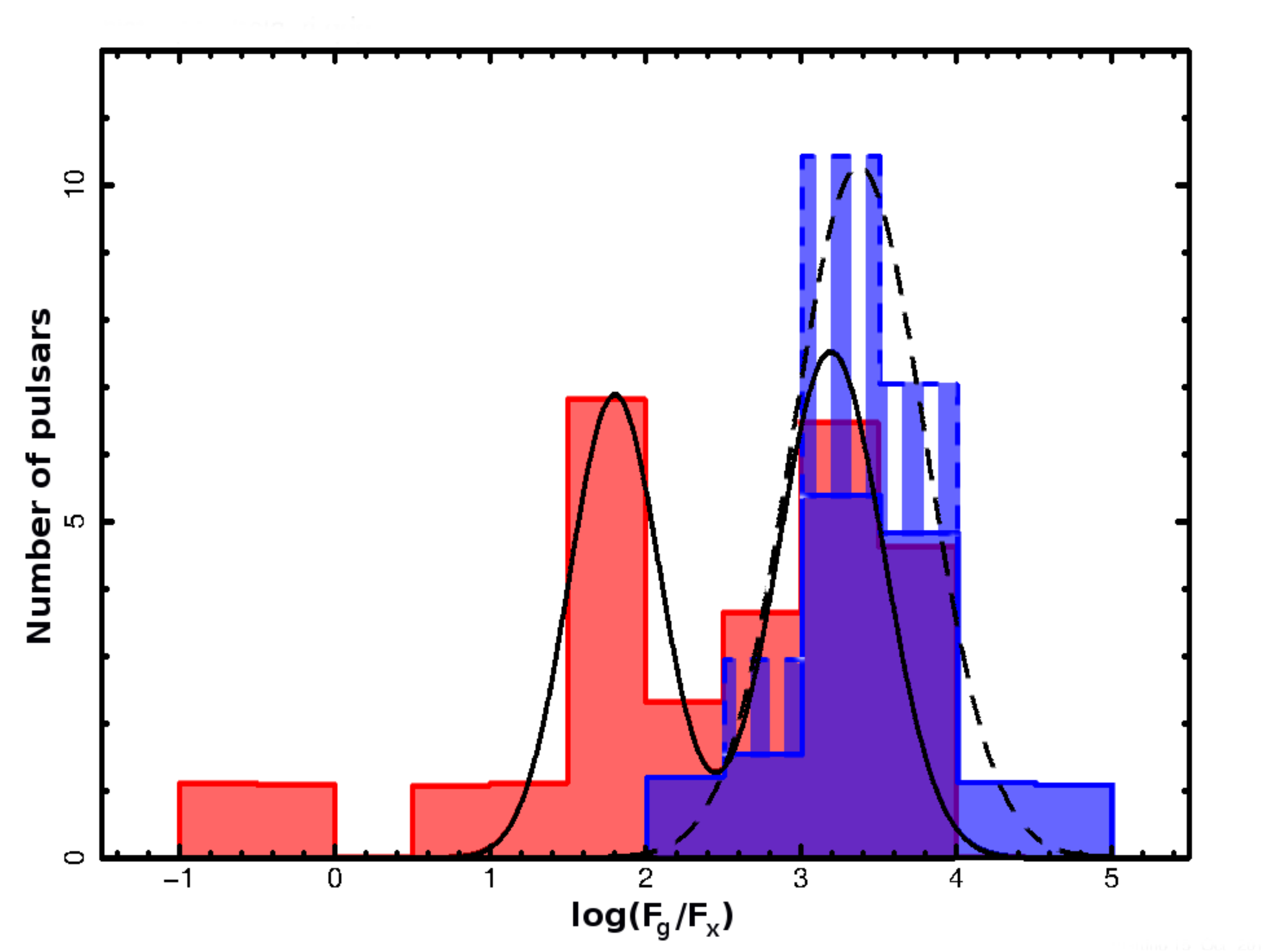}
\protect\caption{{\footnotesize Histogram of the logarithm of the $\gamma$--to--X-ray flux ratio (F$_{\gamma}$/F$_{\rm X}$) of LAT pulsars with high-confidence X-ray detections, as defined in Figure 3 of \citet{mar11}. Histograms for RL and RQ pulsars are shown in red and blue, respectively. The increment to the radio-quiet pulsar histogram for the eight pulsars that we detected in the X rays for the first time is shown in dashed blue and is added to the histogram of previously-known radio-quiet pulsars. For these pulsars, we used the results of the PL fits to compute the F$_X$ values. The continuous and dashed lines report the best gaussian fit for the RL and RQ distributions, respectively.}
\label{histo}}
\end{figure*}

In addition to those obtained from our observations, we collected $\gamma$- and X-ray best-fit spectra from the 2PC and
papers published afterwards (PSR\, J1357$-$6429, \citet{cha12}; J1741$-$2054, \citet{mar14a}; J1813$-$1246, \citet{mar14b};
J0357+3205, \citet{mar13}; J2055+2539, \citet{mar15}). Figure \ref{histo} reports the histogram of the 
Log(F$_{\gamma}$/F$_X$) values for all pulsars, both RL and RQ, detected in the X-rays and with a non-thermal
X-ray spectrum. Improving on the method used in \citet{mar11,mar12}, each pulsar has been represented with an
asymmetric parabola, in order to account for the asymmetric errors on the X- and $\gamma$-ray fluxes. We separately
highlighted in Figure \ref{histo} the Log(F$_{\gamma}$/F$_X$) values of the eight pulsars with the newly discovered
X-ray counterparts (Table 2), assuming a non-thermal X-ray spectrum for all of them. We fitted both RQ and RL pulsar
distributions with single and double gaussian models, taking into account Poissonian errors.

We confirmed that RQ pulsars are intrinsically fainter in the X rays than the RL ones, with an average
Log(F$_{\gamma}$/F$_X$) of 3.38$\pm$0.10 ($1\sigma$ error): at mean, radio-loud pulsars have F$_{\gamma}$/F$_X$ values
an order of magnitude lower (see later).
The distribution of the F$_{\gamma}$/F$_X$ values for RQ
pulsars is well-fitted by a gaussian, with a null hypothesis probability of 0.62, three degrees of freedom (dof),
featuring a sharp peak (standard deviation of 0.43$\pm$0.09). This indicates very similar X- and $\gamma$-ray
emissions among the members of this family. The F$_{\gamma}$/F$_X$ values computed for the eight RQ pulsars with
newly discovered counterparts (Table 2) are distributed around the peak, significantly increasing our statistics. We note that for four of
the eight pulsars the X-ray spectrum could also be fitted by a single BB model, which would result in an even higher
value of the corresponding F$_{\gamma}$/F$_X$.

As noted from Figure \ref{histo}, the F$_{\gamma}$/F$_X$ values of RL pulsars are characterized by a
more structured distribution. While a single gaussian could fit the distribution, with an average Log(F$_{\gamma}$/F$_X$)
of 2.24$\pm$0.32, the null hypothesis probability results quite low (0.05, eight dof) and the distribution is much wider,
with a standard deviation of 0.72$\pm$0.11. By fitting the distribution with two gaussians, we obtain two peaks at
1.81$\pm$0.11 (standard deviation of 0.29$\pm$0.09) and 3.20$\pm$0.14 (standard deviation of 0.34$\pm$0.05), with the
better null hypothesis probability of 0.38 (five dof). An f-test \citep{bev69} shows that the probability for a
chance improvement is 0.03, not enough to exclude the single-gaussian fit.

Interestingly enough, the second peak of the F$_{\gamma}$/F$_X$ distribution of RL pulsars would overlap the peak of
the corresponding distribution for the RQ pulsars. If statistically confirmed, this might suggest that the separation
between RL and RQ pulsars in the F$_{\gamma}$/F$_X$ space might not be as clear as it was originally thought. At least
partially, the overlap between the two peaks might be attributed to the subtle, and somehow arbitrary, distinction between radio-faint and radio-quiet pulsars. 
Indeed, some RL pulsars with the highest F$_{\gamma}$/F$_X$ values (e.g., J1741$-$2054 and J1907+0602, see Figure 3 in the 2PC) are radio-faint,
whereas some RQ pulsars with the lowest F$_{\gamma}$/F$_X$ values, e.g. J1813$-$1246, are very distant and absorbed. Thus,
they would be undetected if they had a radio luminosity comparable to those of J1741$-$2054 and J1907+0602. In any case,
we found no clear correlation between the radio flux and the X- and $\gamma$-ray fluxes, which might have 
suggested that, e.g. radio-faint pulsars have higher F$_{\gamma}$/F$_X$ values than the radio-bright ones.

\begin{figure*}
\centering
\includegraphics[angle=0,width=15cm]{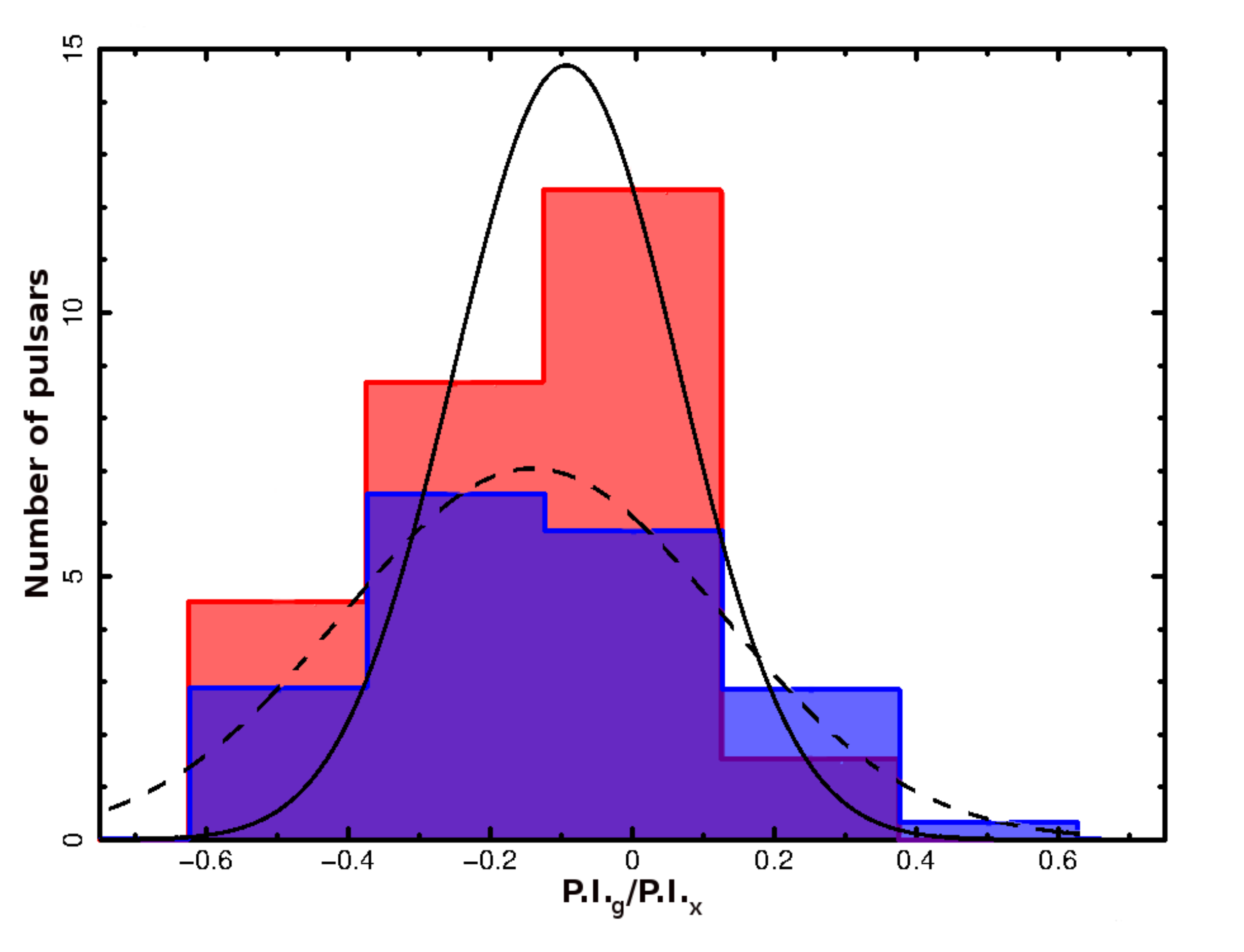}
\protect\caption{{\footnotesize Histogram of the best fitted $\gamma$--to--X-ray photon index ratio ($\Gamma_{\gamma}$/$\Gamma_{\rm X}$) of LAT pulsars with high-confidence X-ray detections, as defined in Figure 3 of \citet{mar11}. Histograms for RL and RQ pulsars are shown in red and blue, respectively. The pulsars for which the photon index has not been fitted are not included. The continuous and dashed lines report the best gaussian fit for the RL and RQ distributions, respectively.}
\label{histo2}}
\end{figure*}

In a similar way, we built the histogram of the photon indices ratio, $\Gamma_{\gamma}/\Gamma_X$, for
RL and RQ pulsars families (Figure \ref{histo2}).
In both cases, the distributions are well fitted by a gaussian
(null hypothesis probabilities of 0.16 and 0.62, 2 dof for RL and RQ pulsars, respectively).
Interestingly enough, the peaks of both distributions occur at very similar values of the $\Gamma_{\gamma}/\Gamma_X$ ratio,
-0.09$\pm$0.03 and -0.14$\pm$0.08 for the RL and RQ pulsars, respectively. This suggests that the peculiar
distribution seen in the F$_{\gamma}$/F$_X$ histogram (Figure \ref{histo}) is not produced by an intrinsic difference
between the spectral slopes of the two families. In the following, we explore some implications of the observed
bimodal F$_{\gamma}$/F$_{X}$ distribution for RQ and RL pulsars on pulsar emission models.

Pulsar magnetospheric radiation is highly anisotropic and a complex antenna pattern (i.e., the direction dependence of the
emitting power) results as the neutron
star rotates and its beam sweeps the sky. In a specific energy range, different emission models expect different
antenna patterns -- the emission mechanism, geometry and luminosity depend on the magnetic field configuration,
inclination angle and intensity, as well as on the pulsar period, but the physics ruling such mechanisms is not
yet understood. For a complete discussion see e.g. \citet{bai10}.
Constraining the multi-wavelength antenna patterns as a function of different pulsar properties
would yield crucial clues to understand pulsar magnetospheres. 

The observed flux in a given energy range is a phase-averaged cut through the antenna pattern for the Earth line of sight.
A beaming factor $f_{\Omega}$ is usually defined as the ratio of the observed flux to the average
flux over $4\pi$ sr \citep[e.g.][]{wat09}.
In the radio range, the antenna pattern is usually described with the $"$cone plus core$"$ heuristic model
\citep[e.g.][]{har07}. According to this model, then, the radio emission is centered on the magnetic axis
(in a co-rotating frame) and radio loudness (or, quietness) is merely a fortuitous effect, set by our line of
sight intercepting (or, missing) the radio antenna pattern of a pulsar, as a result of a small (or, high) magnetic
impact angle. In other words, RQ pulsars are simply those that are seen under a viewing angle substantially different
from their magnetic inclination angle -- the beaming factor for such lines of sight being essentially zero.

In the $\gamma$-ray range, recent works based on data collected by the {\em Fermi}-LAT \citep[see e.g.][]{bai10,pie14} showed very complex antenna patterns,
best described by outer magnetospheric models such as the Outer Gap \citep{rom96} or the Two-pole Caustic model \citep{mus04}. According to
such models, RQ pulsars should not have, on average, a different $\gamma$-ray beaming factor with respect to RL
ones (e.g. \citet{pie14} -- indeed, beaming factors for RQ pulsars are expected to be more dispersed
than for RL pulsars). On the other hand, although distance estimates for RQ pulsars are highly uncertain, there is
no evidence for a larger luminosity (nor for a larger 
spin-down conversion efficiency n$_{\gamma} \equiv $L$_{\gamma}/\dot{E}$, as defined in \citet{abd10}) with respect to the RL ones, although some selection bias, related to highly
pulsed $\gamma$-ray signal and little timing noise
easing blind periodicity searches in $\gamma$-ray data, is certainly affecting our view of RQ pulsars.

In the above picture, it is tempting to link the different F$_{\gamma}$/F$_{X}$ distributions for RQ and RL pulsars
to differences in their X-ray emission properties,
which could allow us to set first constraints on the poorly known X-ray antenna pattern. 
For the RQ pulsars (seen with large magnetic impact angle) the X-ray emission could possibly come from the outer
magnetosphere, with an antenna pattern (and a beaming factor) possibly similar to the $\gamma$-ray one. This could explain
the narrow F$_{\gamma}$/F$_{X}$ distribution for RQ pulsars in spite of their expected large dispersion in $\gamma$-ray
beaming factors. Focusing on the RL pulsars, about half of the sample has Log$(F_{\gamma}/F_{\rm X})\sim1.8$, much lower than the
average Log$(F_{\gamma}/F_{\rm X})\sim3.4$
for the RQ pulsars. This could suggest the existence of a luminous X-ray emission component that can only be seen for
small magnetic impact angles, and thus
possibly centred on the magnetic axis. 
\citet{mar14b} explained the peculiar multi-wavelength
behavior of the RQ pulsar J1813$-$1246 by modeling its X-ray emission using a polar cap model \citep{dyk04}:
a low-altitude cone beam with peak emission just inside the polar cap rim. The radio beam, centred on the X-ray cone,
is missed by just a few degrees: a slightly different line of sight would have made J1813$-$1246 a bright radio pulsar.
Such an X-ray polar cap emission component should display a large variability in luminosity and/or beaming factor within
the RL pulsar sample in order to account for the large dispersion in F$_{\gamma}/$F$_{\rm X}$. Such a variability could be driven
by other pulsar properties such as the spin-down-luminosity and/or the magnetic field configuration/intensity, as well as
their evolution as a function of time.

A major step forward in our understanding of the overall multi-wavelength emission geometry of pulsars could be obtained
by performing simultaneous fitting of their multi-wavelength light curves. Recently, \citet{pie14}
have jointly fit $\gamma$-ray and radio light curves with simulated $\gamma$-ray and radio emission patterns. This allowed
them to investigate some relations between observable characteristics and intrinsic pulsar parameters, encouraging the
creation and testing of new models \citep[see e.g.][]{kal12,li12}. Also owing to the increasing number of RQ pulsars, the 
above analysis should be extended to include X-ray light curves, taking into account both thermal and non-thermal
pulsed emission. Simultaneous fitting and phase-resolved spectral analysis of multi-wavelength emission of pulsars
will allow us to test different emission models and to build new models for the overall emission,
also better explaining the results presented in this paper.
Unfortunately, of the 28 RQ LAT pulsars now detected in the X-rays, only seven have been observed for
a sufficiently long integration time to detect X-ray pulsations.
Similarly, of the 30 RL LAT pulsars with an X-ray counterpart, only 15 are known to pulsate in the X-ray
band. Therefore, an important contribution to the theoretical analysis would come from the detection of
X-ray pulsations for most of the brightest pulsars through deep follow-up \xmm\ and \chan\ observations.
While a number of much brighter RQ and RL pulsars are better suited for such a future study, deep observations
of \xmm\ , the best in-flight telescope to perform such a search, could allow us to detect pulsations from our eight pulsars.
For instance, if we assume a Lorentzian light curve with a duty cycle of 0.1, with an entire orbit of XMM observation (130 ks)
we would detect pulsations at 5$\sigma$ in case of a pulsed fraction of $\sim$40\% for the brightest of our eight pulsars to
100\% for the weakest one.

\acknowledgments
We thank Andrea Belfiore ("Mario") for the useful discussions on the $\gamma$-ray timing positions of LAT pulsars.
The research leading to these results has received funding from the European Commission Seventh
Framework Programme (FP7/2007-2013) under grant agreement n. 267251. This work was supported by the ASI-INAF contract I/037/12/0,
art.22 L.240/2010 for the project $''$Calibrazione ed Analisi del satallite NuSTAR$"$.
Support for this work was provided by the National Aeronautics and Space Administration through Chandra Award Numbers GO2-13093X and 
GO3-14075X issued by the Chandra X-ray Observatory Center, which is operated by the Smithsonian Astrophysical Observatory
for and on behalf of the National Aeronautics Space Administration under contract NAS8-03060.

{\it Facilities:} \facility{XMM-Newton}, \facility{Chandra}, \facility{VISTA}, \facility{UKIRT}

\begin{table*}
\begin{center}
\begin{tabular}{cllcccccrl}
\tableline\tableline
Name & RA(J2000) & Dec(J2000) & P & $\dot{P}$ & $\dot{E}$ & $\tau_c$ & D$_{\gamma}$\\
 & $^{\rm h}~^{\rm m}~ ^{\rm s}$ & $^{\circ} ~ ^{'} ~ ^{''}$ & ms & 10$^{-14}$ s s$^{-1}$ & 10$^{34}$ erg s$^{-1}$ & ky & kpc\\
 \tableline\tableline
J0734$-$1559 & 07 34 45.7 & $-$15 59 18.8 & 155 & 1.25 & 13.2 & 200 & 1.3\\
J1429$-$5911 & 14 29 58.6 & $-$59 11 36.0 & 116 & 3.05 & 77.4 & 60 & 1.7\\
J1838$-$0537 & 18 38 56.0 & $-$05 37 09.0 & 146 & 46.5 & 593 & 5 & 1.8\\
J1846+0919  & 18 46 25.9 & +09 19 45.7  & 226 & 0.99 & 3.4 & 360 & 1.4\\
J1957+5033  & 19 57 38.4 & +50 33 21.4  & 375 & 0.68 & 0.5 & 870 & 0.8\\
J2028+3332  & 20 28 19.9 & +33 32 4.2  & 177 & 0.49 & 3.5 & 570 & 0.9\\
J2030+4415  & 20 30 51.4 & +44 15 38.8  & 227 & 0.65 & 2.2 & 550 & 0.8\\
J2139+4716  & 21 39 55.9 & +47 16 13.0  & 283 & 0.18 & 0.3 & 2500 & 0.8\\ 
\tableline\tableline
\end{tabular}
\caption{Characteristics of the $\gamma$-ray pulsars discussed in this work. Here, we list the name, $\gamma$-ray timing
coordinates, period ($P$), period derivative ($\dot{P}$),
energetics ($\dot{E}$), characteristic age (($\tau$), and pseudo-distance (D$_{\gamma}$), respectively.
These values are taken from \citet{abd13}. Pulsars positions are taken from the last LAT $\gamma$-ray Pulsar Timing Models,
where possible, and from \citet{ple12a} for PSR\, J1838$-$0537.}
\end{center}
\end{table*}

\begin{landscape}
\begin{table*} 
\begin{center}
\begin{tabular}{lcccccccc}
\tableline\tableline
Name & N$_{\rm H}$ & $\Gamma_{\rm X}$ & F$^{\rm PL}_{\rm X}$ & F$_{\gamma }$/F$_{\rm X}$ & kT & R & F$^{\rm BB}_{\rm X}$ & F$_{\gamma }$/F$_{\rm X}$\\
 & 10$^{21}$ cm$^{-2}$ & & 10$^{-14}$ erg cm$^{-2}$s$^{-1}$ & & eV & m & 10$^{-14}$ erg cm$^{-2}$s$^{-1}$ & \\
 \tableline\tableline
J0734$-$1559 & 2 & 2 & 1.5$\pm$0.6 & 3700$\pm$800 & 200 & 76$_{-11}^{+16}$ & 1.2$\pm$0.4 & $>$4700\\
J1429$-$5911 & 3 & -0.1$\pm$0.7 & 3.3$\pm$1.8 & 2400$\pm$400 & - & - & - & -\\
J1838$-$0537 & $27^{+44}_{-21}$ & 0.8$^{+1.1}_{-0.9}$ & 7.2$\pm$0.9 & 2600$\pm$200 & - & - & - & -\\
J1846+0919  & 2 & 2 & 0.9$\pm$0.2 & 2800$\pm$700 & 200 & 45$_{-13}^{+12}$ & 0.5$\pm$0.3 & $>$5300\\
J1957+5033  & $<$0.25 & 2.1$\pm$0.3 & 3.0$\pm$0.5 & 870$\pm$70 & - & - & - & -\\
J2028+3332  & 2 & 2 & 0.5$\pm$0.3 & 10900$\pm$4700 & 200 & 34$_{-8}^{+12}$ & 0.2$\pm$0.1 & $>$26400\\
J2030+4415  & $0.61^{+1.54}_{-0.61}$ & 2.4$^{+0.8}_{-0.6}$& 2.1$\pm$0.6 & 2800$\pm$400 & - & - & - & -\\
J2139+4716  & 1 & 2 & 0.5$\pm$0.3 & 4900$\pm$1300 & 200 & 33$\pm$9 & 0.3$\pm$0.1 & $>$8500\\ 
\tableline\tableline
\end{tabular}
\caption{Summary of the X-ray spectral parameters for the pulsars discussed in this work, as described in Section 3.
Here, we list the name, best fit column density, photon index, flux and gamma-to-X flux ratio in the case of non-thermal emission
and temperature, radius, flux and lower limit gamma-to-X flux ratio in case of thermal emission, where acceptable.
All quoted errors are at the 90\% c.l., with the exception of the F$_{\gamma }$/F$_{\rm X}$ for which the 1$\sigma$ error is reported.}
\end{center}
\end{table*}
\end{landscape}

\clearpage

\end{document}